\documentclass[aps,twocolumn,preprintnumbers,amsmath,amssymb,floats,citeautoscript]{revtex4-1}
\usepackage{graphicx}
\usepackage{dcolumn}
\usepackage{bm}
\usepackage{times}

\newcommand{\vect}[1]
{{\mbox{\boldmath $#1$}}}

\begin{document}


\title{Collective excitation of a short-range charge ordering in $\theta$-$\mbox{(BEDT-TTF)}_2\mbox{CsZn(SCN)}_4$}

\author{K.~Hashimoto$^{1,2}$}
\author{S.\,C.~Zhan$^{1}$}
\author{R.~Kobayashi$^{1}$}
\author{S.~Iguchi$^{1,2}$}
\author{N.~Yoneyama$^{2,3}$}
\author{T.~Moriwaki$^{4}$}
\author{Y.~Ikemoto$^{4}$}
\author{T.~Sasaki$^{1,2}$}

\affiliation{$^1$Institute for Materials Research, Tohoku University, Aoba-ku, Sendai 980-8577, Japan\\
$^2$CREST, Japan Science and Technology Agency, Tokyo 102-0075, Japan\\
$^3$Department of Education, Interdisciplinary Graduate School of Medicine and Engineering,\\
University of Yamanashi, Kohu, Yamanashi 400-8511, Japan\\
$^4$SPring-8, Japan Synchrotron Radiation Research Institute, Sayo, Hyogo 679-5198, Japan}

\date{\today}



\begin{abstract}
We find a characteristic low-energy peak structure located in the range of 100--300 cm$^{-1}$ in the optical conductivity spectra of a quasi-two-dimensional organic compound with a triangular lattice, ${\theta}$-$\mbox{(BEDT-TTF)}_2\mbox{CsZn(SCN)}_4$, in which two different types of short-range charge orderings (COs) coexist. Upon lowering the temperature, the low-energy peak becomes significant and shifts to much lower frequencies only for the polarization of ${\vect{E} \parallel \vect{a}}$, in contrast to the other broad electronic bands in the mid-infrared region. On introducing disorder, the low-energy peak is strongly suppressed in comparison with the broad electronic bands. This result indicates that the low-energy peak is attributed to a collective excitation that originates from the short-range CO with a relatively long-period $3\times3$ pattern. The present results shed light on the understanding of the low-energy excitation in the glassy electronic state, where the charge degrees of freedom remain at low temperatures.

\end{abstract}

\maketitle

Charge ordering (CO) has become an important physical concept for understanding insulating ground states of quarter-filled electron systems \cite{Takahashi06,Seo06}. In conventional one-dimensional (1D) systems, nesting of the 1D Fermi surface, combined with the electron-phonon interaction, induces a Peierls instability, which eventually leads to a charge-density-wave order. Contrastingly, in strongly correlated electron systems such as organic conductors and transition-metal oxides, Wigner-type CO often emerges; in this type of CO, the off-site Coulomb interaction plays an important role in the formation of localized electrons in real space, mostly leading to long-range CO. Recently, however, in geometrically frustrated systems, insulating ground states without long-range CO have been observed \cite{Kagawa13}. Geometrical frustration can suppress the tendency towards long-range CO, since competition among various CO patterns due to geometrical frustration prevents a specific charge configuration, which is analogous to geometrically frustrated spin systems, such as quantum spin liquids or spin glasses. Thus, geometrical frustration may lead to exotic electronic states, such as a charge-glass state or a quantum melting of CO \cite{Merino05,Kaneko06,Watanabe06}, in which charge degrees of freedom remain at low temperatures. 

Quasi-two-dimensional (quasi-2D) organic compounds with a triangular lattice, $\theta$-$\mbox{(BEDT-TTF)}_2X$, where BEDT-TTF denotes the donor molecule bis(ethylenedithio)-tetrathiafuvalen and $X$ represents a monovalent anion, have been extensively studied because of the occurrence of a CO metal-insulator transition \cite{Kagawa13,Merino05,Kaneko06,Watanabe06,H_Mori98,Sawano05,Yamaguchi06,Chiba08,Clay02,Mori03,Hotta06,Kuroki06,Udagawa07,Nishimoto08}. The crystal structure consists of an alternating stack of BEDT-TTF and $X$ layers, and the charge transfer between these layers leads to a quarter-filled hole band system. For the $\theta$-type salts, the BEDT-TTF molecules form a 2D conducting plane with a triangular lattice, where the ratio of the nearest-neighbor Coulomb repulsions $V_c$ and $V_p$ is close to unity \cite{Mori03} (see the inset of Fig.\:1(c)). The ground state of $\theta$-$\mbox{(BEDT-TTF)}_2X$, ranging from charge ordered insulating states to a metallic (superconducting) state, can be tuned using the dihedral angle $\phi$ between BEDT-TTF molecules, which depends on the anion $X$ or pressure \cite{H_Mori98}. It has been shown that $\phi$ alters the intermolecular transfer integrals $t_c$ and $t_p$ (with a smaller $\phi$ yielding a smaller ratio of $t_c$ to $t_p$), and that the CO transition temperature $T_{\rm{co}}$ is lowered with decreasing $\phi$. Indeed, $\theta$-$\mbox{(BEDT-TTF)}_2\mbox{RbZn(SCN)}_4$ (abbreviated as $\theta$-RbZn) with $\phi = 111^{\circ}$ ($t_c/t_p \sim 0.4$) is a well-known long-range CO system ($T_{\rm{co}} = 190$ K) with a horizontal-stripe structure along the $c$-axis \cite{Watanabe03,Watanabe04}. In contrast, $\theta$-$\mbox{(BEDT-TTF)}_2\mbox{I}_3$ with $\phi = 100^{\circ}$ exhibits superconductivity at 3.6 K \cite{Kobayashi86}. What is even more intriguing in the $\theta$-type salts is that $\theta$-$\mbox{(BEDT-TTF)}_2\mbox{CsZn(SCN)}_4$ ($\theta$-CsZn) with $\phi = 104^{\circ}$ ($t_c/t_p \sim 0.1$) located on the phase boundary exhibits no clear long-range CO but rather the coexistence of two kinds of short-range COs with modulation wave vectors $\vect{q}_{1\rm{d}}=(2/3, k, 1/3)$ and $\vect{q}_{2\rm{d}}=(0, k, 1/2)$ \cite{Watanabe99,Nogami99} ; the charge modulation with $\vect{q}_{1\rm{d}}$ corresponds to a $3\times3$ CO with respect to the $a$-$c$ unit cell, while the modulation with $\vect{q}_{2\rm{d}}$ is the same as the horizontal-stripe CO observed in $\theta$-RbZn.
Theoretically, it has been pointed out that the competition among various Wigner-type COs due to geometrical frustration \cite{Merino05,Kaneko06,Watanabe06} can cause such a glassy CO state, but a consensus for the origin of the CO state in $\theta$-CsZn is still lacking. Therefore, it is highly desirable to further investigate details of the charge dynamics.

Here, we report optical conductivity measurements of $\theta$-CsZn. We have found that the optical conductivity spectra can be described by three characteristic structures, one of which, located at 100 cm$^{-1}$, is very sensitive to molecular defects introduced by x-ray irradiation. Considering the long-distance Coulomb interactions, we discuss the possibility of a collective excitation derived from the short-range $3\times3$ CO as the origin of the low-energy excitation. 


\begin{figure}[tb]
\centering
\includegraphics[width=1.02\linewidth]{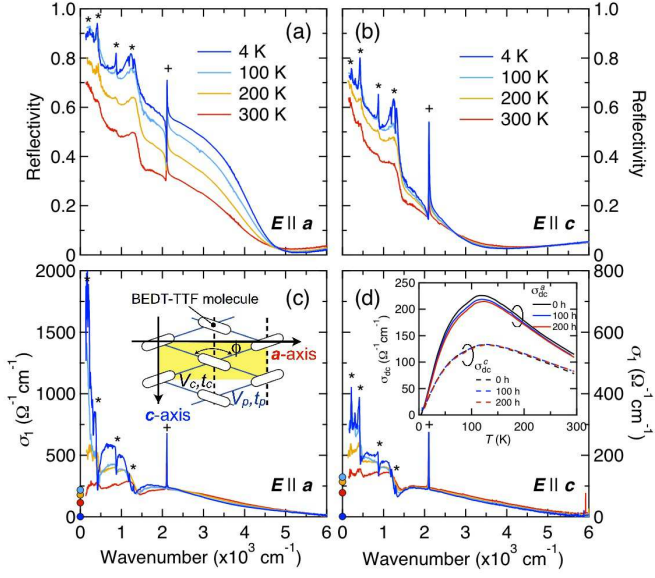}
\caption{ (Color online). Optical reflectivity spectra of $\theta$-CsZn for (a) $\vect{E} \parallel \vect{a}$ and (b) $\vect{E} \parallel \vect{c}$ at various temperatures. Optical conductivity spectra of $\theta$-CsZn for (c) $\vect{E} \parallel \vect{a}$ and (d) $\vect{E} \parallel \vect{c}$. The circles at $\omega = 0$ denote the dc conductivity at each temperature. The asterisks indicate the intramolecular vibrations of the BEDT-TTF molecule, and the cross denotes the CN stretching mode. The inset in (c) shows a schematic of the arrangement of BEDT-TTF molecules for $\theta$-$\mbox{(BEDT-TTF)}_2X$. The shaded area denotes the unit cell. $\phi$ is the dihedral angle between BEDT-TTF molecules. 
The intermolecular transfer integrals (Coulomb repulsions) are labeled as $t_p$ ($V_p$) and $t_c$ ($V_c$). The inset of (d) shows the temperature dependence of the in-plane dc conductivity $\sigma_{\rm{dc}}^{a,c}(T)$ for several irradiation times. 
} \label{Optical}
\end{figure}

Single crystals of $\theta$-CsZn were grown through an electrochemical oxidation method \cite{H_Mori98}. Polarized reflectivity measurements were performed with a Fourier transform microscope spectrometer in the range of 100--45000 cm$^{-1}$. In the far-infrared region (100--650 cm$^{-1}$), synchrotron radiation light at BL43IR in SPring-8 was used. The reflectivity was determined by comparison with a gold thin film evaporated on the sample surface. 
The optical conductivity was calculated through a Kramers-Kronig (KK) transformation. The reflectivity in the range of 8000--45000 cm$^{-1}$ was measured at room temperature. 
Above 45000 cm$^{-1}$ the standard $\omega^{-4}$ extrapolation was used. At low frequencies, both constant and linear extrapolations were applied. We have confirmed that there is little difference between the two extrapolation procedures in the measured region. The in-plane dc conductivity was measured in the linear $I$-$V$ region. 
For introducing molecular defects, the crystals were irradiated by x-rays at room temperature using an unfiltered tungsten target at 40 kV and 20 mA. The dose rate was estimated to be 0.5 MGy/h \cite{Sasaki08}. The temperature dependence of the resistivity and reflectivity were measured after each step of the irradiation.


Figures\:1(a) and 1(b) show the optical reflectivity spectra at various temperatures for the polarizations of $\vect{E} \parallel \vect{a}$ and $\vect{E} \parallel \vect{c}$, respectively. Upon lowering the temperature, the reflectivity increases for both directions. Several vibrational features can be seen at 220, 400, 880, and 1230 cm$^{-1}$; these are attributed to the totally symmetric $a_g$ vibrational modes of the BEDT-TTF molecule coupled with electronic excitations through the electron-molecular vibration (emv) interaction \cite{Tajima00,Wang03,Suzuki05}. The sharp peak at approximately 2100 cm$^{-1}$ is the CN stretching mode of (SCN)$^{-1}$ in the anion layer. 

Figures\:1(c) and 1(d) display the optical conductivity spectra $\sigma_1(\omega)$ obtained from the KK transformation for $\vect{E} \parallel \vect{a}$ and $\vect{E} \parallel \vect{c}$, respectively. At room temperature, the optical spectra for $\vect{E} \parallel \vect{a}$ and $\vect{E} \parallel \vect{c}$ show similar behavior. The spectral intensity for $\vect{E} \parallel \vect{a}$ is approximately 2.5 times larger than that for $\vect{E} \parallel \vect{c}$, reflecting the difference between the intermolecular transfer integrals in the $a$ and $c$ directions. With decreasing temperature, the spectral weight (SW) below $\sim$1500 cm$^{-1}$ increases for both directions; this behavior is totally different from the behavior observed for slowly cooled $\theta$-RbZn, in which a well-established charge gap is observed at low temperatures \cite{Wang01}. What is more remarkable is that $\sigma_1(\omega)$ in the low-energy region below $\sim$500 cm$^{-1}$ is strongly enhanced by lowering the temperature, especially for $\vect{E} \parallel \vect{a}$. One should note that the enhancement of $\sigma_1(\omega)$ for $\vect{E} \parallel \vect{a}$ differs from a Drude response since the dc conductivity $\sigma_{\rm{dc}}$ decreases to $\sim$1 $\Omega^{-1} \rm{cm}^{-1}$ at 4 K (see the inset of Fig.\:1(d)). Therefore, $\sigma_1(\omega)$ should have a peak structure below $\sim$100 cm$^{-1}$. At low temperatures, in addition to this low-energy peak,  it can be clearly seen that there are two broad bands located at approximately 800 and 2200 cm$^{-1}$ (also see Fig.\:2). Besides, the antisymmetric (or antiresonance) features of the vibrational modes of the BEDT-TTF molecule in the far-infrared region become more noticeable because of the increase of the electronic background at low temperatures. 


To distinguish the electronic and vibrational contributions,  at low temperatures we fit the data to a model with Lorentzian components describing the electronic excitations and Fano components reflecting the intramolecular vibrations:
\begin{eqnarray}
\hat{\sigma}(\omega)
&=&
\hat{\sigma}^{\rm{Lorentz}}+\hat{\sigma}^{\rm{Fano}} \nonumber \\
&=&
\frac{Ne^2}{m_0}\frac{\omega}{i(\omega_0^2-\omega^2)+\gamma\omega}
+
\frac{\sigma_0(q-i)^2\gamma\omega}{i(\omega_0^2-\omega^2)+\gamma\omega},
\label{Sigma_eq}
\end{eqnarray}
where $N$ is the number of charge carriers, $m_0$ is the band mass, $\omega_0$ is the center frequency, $\gamma$ is the linewidth, $\sigma_0$ is the electronic background, and $q$ is the Fano parameter characterizing the degree of line asymmetry.  At higher temperatures, the dc conductivity seems to be not negligible for the analysis. 
Recent THz transmission measurements for $\theta$-CsZn \cite{Shimano} have revealed that an incoherent state, often seen in bad metals, is observed around 30 cm$^{-1}$ at room temperature. This incoherent state can be still seen below 100 K, and evolves to a gapped state below 20 K, which is consistent with the fact that $\sigma_{\rm{dc}}$ obeys the Arrhenius law below 20 K \cite{Suzuki05}. Therefore, the low-energy peak structure observed in the present study differs from the incoherent peak, implying that the finite value of the dc conductivity is not due to the contribution from the low-energy peak, but from the incoherent state in the THz region. In order to consider the contribution from the finite value of $\sigma_1(0)$, we simply added a narrow Drude term $L_{\rm{incoherent}}$ into Eq.\:(1), since the incoherent response with a maximum near the dc limit differs little from a narrow Drude response. As shown in Figs.\:2(a), 2(b), 2(d), and 2(e), the incoherent component rapidly decays with increasing frequency as expected from the small value of $\sigma_1(0)$. Therefore, the Lorentzian terms are not affected by the incoherent state. Thus, the optical spectra are mainly composed of three characteristic Lorentzian curves with center frequencies of 100--300, 800--1000, and 2200--2500 cm$^{-1}$ (referred to as $L_{\rm{low}}$, $L_{\rm{middle}}$, and $L_{\rm{high}}$, respectively). 
As shown in Figs.\:3(b), 3(c), 3(f), and 3(g), the center frequencies of $L_{\rm{middle}}$ and $L_{\rm{high}}$ are located at almost the same position for both polarizations at each temperature, and they shift to lower frequencies by about 10--20\% when the crystals are cooled to 4 K. In contrast, the center frequency of the low-energy peak $L_{\rm{low}}$ shows a different trend with temperature for $\vect{E} \parallel \vect{a}$ and $\vect{E} \parallel \vect{c}$, as shown in Figs.\:3(a) and 3(e); the center frequency for $\vect{E} \parallel \vect{a}$ exhibits a large shift from 300 to 100 cm$^{-1}$ (i.e., a decrease of about 70\%), whereas for $\vect{E} \parallel \vect{c}$ it shows only a slight change. It is noteworthy that the strong temperature dependence of the center frequency of $L_{\rm{low}}$ observed for the polarization of $\vect{E} \parallel \vect{a}$ was not seen in any other quasi-2D organic compounds.

\begin{figure}[tb]
\centering
\includegraphics[width=0.98\linewidth]{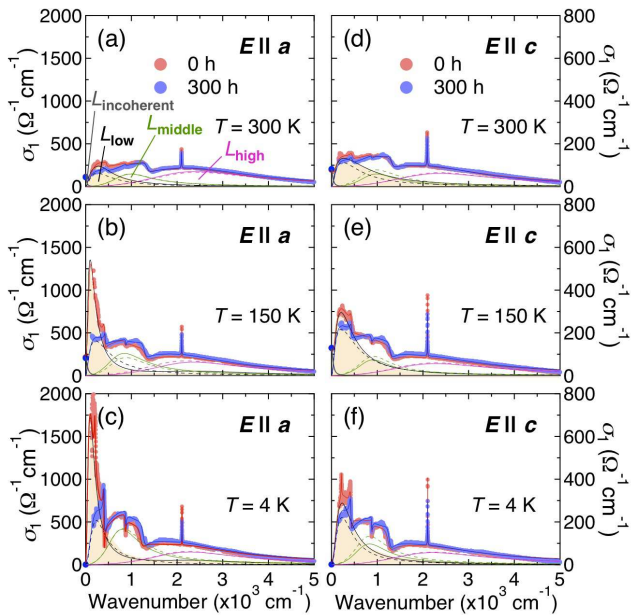}
\caption{ (Color online). (a)--(f) Optical conductivity spectra of $\theta$-CsZn before (red) and after (blue) x-ray irradiation for 300 h at 4, 150, and 300 K for $\vect{E} \parallel \vect{a}$ and $\vect{E} \parallel \vect{c}$. The red and blue circles at $\omega = 0$ indicate the dc conductivity before and after x-ray irradiation, respectively. The red (blue) lines represent the overall fit before (after) x-ray irradiation. The black, green, and magenta solid (dashed) lines correspond to the bands $L_{\rm{low}}$, $L_{\rm{middle}}$, and $L_{\rm{high}}$ before (after) x-ray irradiation, respectively. The gray lines at higher temperatures indicate the contribution from the conductive carries. The Fano components are not shown for clarity.
} \label{Irradiation}
\end{figure}

The broad bands $L_{\rm{middle}}$ and $L_{\rm{high}}$ have often been observed in other quarter-filled organic conductors close to a CO phase such as the $\alpha$- \cite{Dressel03,Drichko06} and $\beta^{"}$-phase \cite{Kaiser10} BEDT-TTF salts, as well as in $\theta$-RbZn above $T_{\rm{co}}$ \cite{Tajima00,Wang01}. Numerical calculations based on the extended Hubbard model that takes into account the nearest-neighbor Coulomb interaction at quarter-filling on a square lattice \cite{Merino03} predict that in a metallic state close to a CO phase, in addition to a Drude peak, two broad electronic bands emerge in the optical spectra owing to CO fluctuations. A transition between Hubbard-like bands induced by the intersite Coulomb repulsion $V$, i.e., a site-to-site transition within the fluctuating CO pattern, gives rise to a broad band in the mid-infrared region of the order of $V$, which corresponds to $L_{\rm{high}}$ \cite{explanation1}. The other band is a ``charge fluctuation band" originating from short-range CO fluctuations, corresponding to $L_{\rm{middle}}$. This band is enhanced as $V$ is increased, but it becomes invisible as $V$ approaches a critical value at which the long-range CO occurs. Indeed, while the broad band around 2500 cm$^{-1}$ in $\theta$-RbZn shifts to a higher energy region below $T_{\rm{co}}$ = 190 K, the band around 1000 cm$^{-1}$ becomes less pronounced below $T_{\rm{co}}$ \cite{Tajima00,Wang01}, indicating that short-range CO fluctuations are important for the formation of $L_{\rm{middle}}$. In the case of $\theta$-CsZn, there is no Drude response as discussed above. Therefore, $L_{\rm{middle}}$ is considered to be due to a transition between the incoherent state observed in the THz region and the Hubbard-like bands. Thus, the above two bands, $L_{\rm{middle}}$ and $L_{\rm{high}}$, can be well understood in the framework of the extended Hubbard model including the nearest-neighbor Coulomb interactions. The striking result of this work is the low-energy peak around 100 cm$^{-1}$ observed only in the polarization of $\vect{E} \parallel \vect{a}$. One should note that although a similar band is observed at about 300 cm$^{-1}$ in $\beta^{"}$-(BEDT-TTF)$_2$SF$_5$CH$_2$CF$_2$SO$_3$, the band appears in both directions and does not show a strong temperature dependence \cite{Kaiser10}.

There are several key features that appear to be clues for understanding the origin of the low-energy peak $L_{\rm{low}}$. In the x-ray diffuse scattering experiments for $\theta$-RbZn above $T_{\rm{co}} = 190$ K, diffuse rods associated with a short-range $3\times4$ CO are observed at $\vect{q}_{1'\rm{d}}=(\pm1/3, k, \pm1/4)$, which are replaced by the long-range charge modulation with $\vect{q}_2$ below $T_{\rm{co}}$ \cite{Watanabe03,Watanabe04}. However, when the crystals are rapidly cooled through the transition temperature, the above short-range $3\times4$ CO remains, and another diffuse rod appears at $\vect{q}_{2\rm{d}}$ \cite{Watanabe04}. Thus, two kinds of short-range COs coexist below 190 K. This is very similar to what is observed in $\theta$-CsZn, in which x-ray diffuse scattering measurements \cite{Nogami99} revealed that two kinds of diffuse rods associated with the short-range $3\times3$ and horizontal COs start to appear at $\vect{q}_{1\rm{d}}$ and $\vect{q}_{2\rm{d}}$ below $\sim$150 K.
Theoretical calculations for the $\theta$-type salts based on the extended Hubbard model considering the nearest-neighbor Coulomb repulsions $V_c$ and $V_p$ \cite{Kaneko06,Watanabe06,Mori03,Hotta06,Kuroki06,Udagawa07,Nishimoto08} suggest that a $c$-axis three-fold CO, which has a three-fold periodicity only in the $c$ direction, is more stable at high temperatures than the horizontal-stripe CO when $V_c \approx V_p$. Because the phase transition between the above two COs is considered to be first order \cite{Mori03}, the high-temperature phase can be frozen when the system is cooled rapidly or $T_{\rm{co}}$ is sufficiently lowered, resulting in a charge-cluster glass state \cite{Kagawa13}. The extended Hubbard model including $V_c$ and $V_p$ captures the important aspects of the CO state in the $\theta$-type salts; however, the $3\times3$ and $3\times4$ COs observed in the experiments cannot be reproduced within the framework of the model containing only $V_c$ and $V_p$. Thus, the importance of the more distant Coulomb interactions has been discussed \cite{Mori03,Kuroki06,Naka13}. Theoretical calculations including the next-nearest-neighbor Coulomb interactions have revealed that the $3\times3$ and $3\times4$ COs are more stabilized than the $c$-axis three-fold CO, indicating that the long-period Coulomb interactions are important for the formation of the $3\times3$ and $3\times4$ COs. The condition that $V_c$ and $V_p$ are comparable owing to geometrical frustration can make the next-nearest-neighbor Coulomb interactions more important for determining a specific configuration among the various CO patterns.

\begin{figure}[tb]
\centering
\includegraphics[width=1.0\linewidth]{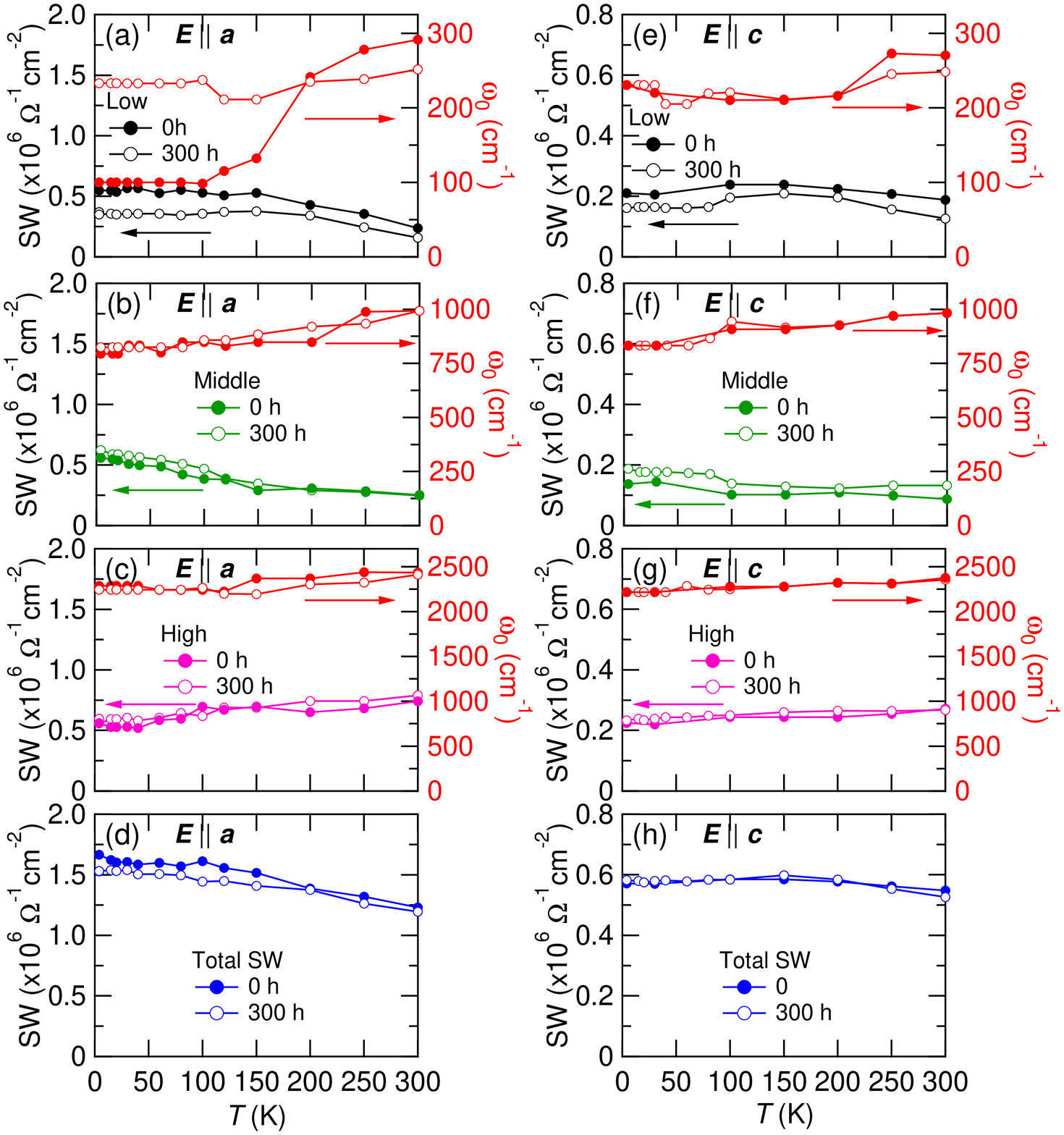}
\caption{ (Color online). Left axis: (a)--(c) and (e)--(g) show the temperature dependence of the SW of $L_{\rm{low}}$, $L_{\rm{middle}}$, and $L_{\rm{high}}$ for, respectively, $\vect{E} \parallel \vect{a}$ and $\vect{E} \parallel \vect{c}$ before (filled circles) and after (open circles) x-ray irradiation for 300 h. Right axis: The same plots for the center frequencies. (d) and (h) show the temperature dependence of the total SW for, respectively, $\vect{E} \parallel \vect{a}$ and $\vect{E} \parallel \vect{c}$ before (filled circles) and after (open circles) x-ray irradiation.
} \label{SW}
\end{figure}

Recent theoretical calculations by Fratini and Merino \cite{Fratini09} based on a 2D Hubbard-Wigner model including the full long-range Coulomb interactions on a square lattice predict the manifestation of a collective excitation in the vicinity of a CO phase. In this situation, the electrons can resonate between nearly-degenerate energy states, which form a collective excitation that can propagate coherently at long distances as a result of a gain in the kinetic energy. As discussed above, the $3\times3$ CO observed in $\theta$-CsZn requires the more-distant (at least next-nearest-neighbor) Coulomb interactions. Therefore, the low-energy peak $L_{\rm{low}}$ in $\theta$-CsZn may be attributed to a collective excitation originating from the $3\times3$ CO. This assignment is consistent with the unidirectional behavior of $L_{\rm{low}}$; i.e., the low-energy peak $L_{\rm{low}}$ becomes sharper and shifts to much lower frequencies only in the polarization of $\vect{E} \parallel \vect{a}$ with decreasing temperature. Recent x-ray diffuse scattering measurements \cite{Kagawa} have revealed that the evolution of the correlation length of the short-range $3\times3$ CO becomes temperature-independent below $\sim$100 K. The development of $L_{\rm{low}}$ for $\vect{E} \parallel \vect{a}$ shows the same trend with temperature, suggesting that the low-energy peak originates from the short-range $3\times3$ CO. The domain size of the $3\times3$ CO is estimated to be $\sim$20 {\AA} at 160 K, which evolves to $\sim$70 {\AA} at 80--100 K and saturates at further low temperatures. Therefore, the development of the low-energy peak for $\vect{E} \parallel \vect{a}$ with cooling is considered to reflect the increase in the domain size, resulting in that the collective excitation can exist more coherently.

To check this, we introduced molecular defects into the crystals through x-ray irradiation; moderate defects would adequately affect the collective excitation. X-ray irradiation effects have been intensively studied for $\kappa$-(BEDT-TTF)$_2X$ \cite{Sasaki08}, and it is well established that nonmagnetic impurities are introduced mainly into the anion layers, because the scattering cross section of the elements in the anion layer is larger than that of the BEDT-TTF molecule \cite {Sasaki08,Yoneyama10}. Thus, the charge carriers in the BEDT-TTF layers are influenced by the random potential modulation in the anion layer. Figure\:1(d) shows the irradiation-time dependence of $\sigma_{\rm{dc}}(T)$ for the $a$ and $c$ directions. The dc conductivity for the $a$ direction, $\sigma_{\rm{dc}}^{a}$, slightly decreases with increasing irradiation time, whereas $\sigma_{\rm{dc}}^{c}$ is nearly unchanged. The variation of $\sigma_{\rm{dc}}^{a}$ is, however, negligibly small when it is compared to the irradiation effects on the optical conductivity $\sigma_1(\omega)$ (see Figs.\:2(a)--2(f)). 

The optical conductivity spectra of $\theta$-CsZn after x-ray irradiation for 300 h for $\vect{E} \parallel \vect{a}$ and $\vect{E} \parallel \vect{c}$ are shown in Figs.\:2(a)--2(f). At room temperature, a slight reduction in $\sigma_1(\omega)$ can be seen in the low-energy region; this reduction becomes more prominent as the temperature is lowered. One can see that the low-energy peak observed in the pristine sample for $\vect{E} \parallel \vect{a}$ below $\sim$150 K is drastically suppressed because of the x-ray irradiation. As shown in Figs.\:3(a) and 3(e), the center frequency becomes nearly temperature-independent for both polarizations, and the SW decreases under x-ray irradiation. As a result, the anisotropic temperature dependence observed in the pristine sample disappeared. This result suggests that the low-energy peak $L_{\rm{low}}$ observed in the pristine sample is a collective excitation sensitive to impurities. The molecular defects introduced by x-ray irradiation could prevent the formation of collective excitation that can propagate coherently over long distances. In contrast, the broad bands $L_{\rm{middle}}$ and $L_{\rm{high}}$ are nearly unchanged under x-ray irradiation, strongly supporting that $L_{\rm{middle}}$ and $L_{\rm{high}}$ originate from the interband transitions depending on fundamental electronic parameters such as the intermolecular transfer integrals and Coulomb interactions. Because the total SW is nearly conserved for both directions before and after x-ray irradiation, the amount of SW reduction of the low-energy peak should be redistributed to higher frequencies.

Finally, we comment on the characteristic temperature dependence of the SW of $L_{\rm{low}}$. As shown in Figs.\:3(d) and 3(h), the total SW for $\vect{E} \parallel \vect{a}$ increases with decreasing temperature and saturates below 100 K, whereas for $\vect{E} \parallel \vect{c}$ it does not change so much with temperature. A similar anisotropic temperature dependence of the total SW has been observed in the $\alpha$-phase BEDT-TTF salts \cite{Drichko06}. The anisotropic variation of the total SW reflects the anisotropic temperature dependence of the intermolecular transfer integrals along with the thermal contraction of the lattice constants; only the transfer integral in the $a$ direction increases upon cooling \cite{Ono97}. Therefore, the development of the low-energy peak for $\vect{E} \parallel \vect{a}$ is considered to be due to the increase of the transfer integral in the $a$ direction; i.e., the reduction in $V/t$ gives rise to the manifestation of the collective excitation. The fact that the spectral weight of the low-energy peak already exists at room temperature for both polarizations implies that the collective excitation which can propagate some distances already exists at room temperature. This is consistent with the $^{13}$C-NMR experiment \cite{Chiba08}, in which extremely slow charge fluctuations associated with the $3\times3$ CO have been already observed at room temperature. 


In summary, we have performed optical conductivity measurements of $\theta$-CsZn before and after x-ray irradiation. Although the two broad bands in the mid-infrared region are relatively insensitive to molecular defects, the low-energy peak around 100 cm$^{-1}$ observed in the pristine sample for $\vect{E} \parallel \vect{a}$ is drastically suppressed because of x-ray irradiation. This result indicates that the low-energy peak originates from a collective excitation derived from the short-range $3\times3$ CO, in which more-distant Coulomb interactions become important owing to geometrical frustration. The present results should be an important input for understanding of the low-energy excitation in the glassy electronic state.

We thank M. Naka, H. Seo, S. Ishihara, M. Watanabe, S. Iwai and I. Terasaki for fruitful discussions. Synchrotron radiation measurements were performed at SPring-8 with the approval of JASRI (2011B1287, 2012A1096, and 2012B1087). This work is supported by Grants-in-Aid for Scientific Research from MEXT (No. 23110702) and JSPS (Nos. 24540357 and 25287080), Japan.



\begin{thebibliography}{99}



\bibitem{Takahashi06}
T. Takahashi, Y. Nogami, and K. Yakushi, 
{J. Phys. Soc. Jpn.} {\bf 75}, 051008 (2006).  

\bibitem{Seo06}
H. Seo, J. Merino, H. Yoshioka, and M. Ogata,
{J. Phys. Soc. Jpn.} {\bf 75}, 051009 (2006).  

\bibitem{Kagawa13}
F. Kagawa, T. Sato, K. Miyagawa, K. Kanoda, Y. Tokura, K. Kobayashi, R. Kumai, and Y. Murakami,
{Nat. Phys.} {\bf 9}, 419, (2013).


\bibitem{Merino05}
J. Merino, H. Seo, and M. Ogata,
{Phys. Rev. B} {\bf 71}, 125111 (2005). 

\bibitem{Kaneko06}
M. Kaneko and M. Ogata,
{J. Phys. Soc. Jpn.} {\bf 75}, 014710 (2006).

\bibitem{Watanabe06}
H. Watanabe and M. Ogata,
{J. Phys. Soc. Jpn.} {\bf 75}, 063702 (2006).

\bibitem{H_Mori98}
H. Mori, S. Tanaka, and T. Mori,
{Phys. Rev. B} {\bf 57}, 12023 (1998).

\bibitem{Sawano05}
F. Sawano, I. Terasaki, H. Mori, T. Mori, M. Watanabe, N. Ikeda, Y. Nogami, and Y. Noda,
{Nature} {\bf 437}, 522 (2005).

\bibitem{Yamaguchi06}
Y. Takahide, T. Konoike, K. Enomoto, M. Nishimura, T. Terashima, S. Uji, and H.\,M. Yamamoto,
{Phys. Rev. Lett.} {\bf 96}, 136602 (2006).

\bibitem{Chiba08}
R. Chiba, K. Hiraki, T. Takahashi, H. M. Yamamoto, and T. Nakamura,
{Phys. Rev. B} {\bf 77}, 115113 (2008).

\bibitem{Clay02}
R.\,T. Clay, S. Mazumdar, and D.\,K. Campbell,
{J. Phys. Soc. Jpn.} {\bf 71}, 1816 (2002). 

\bibitem{Mori03}
T. Mori,
{J. Phys. Soc. Jpn.} {\bf 72}, 1469 (2003).

\bibitem{Hotta06}
C. Hotta and N. Furukawa,
{Phys. Rev. B} {\bf 74}, 193107 (2006).

\bibitem{Kuroki06}
K. Kuroki,
{J. Phys. Soc. Jpn.} {\bf 75}, 114716 (2006).

\bibitem{Udagawa07}
M. Udagawa and Y. Motome,
{Phys. Rev. Lett.} {\bf 98}, 206405 (2007).

\bibitem{Nishimoto08}
S. Nishimoto, M. Shingai, and Y. Ohta,
{Phys. Rev. B} {\bf 78}, 035113 (2008).

\bibitem{Watanabe03}
M. Watanabe, Y. Noda, Y. Nogami, and H. Mori,
{Synth. Met.} {\bf 135-136}, 665 (2003).

\bibitem{Watanabe04}
M. Watanabe, Y. Noda, Y. Nogami, and H. Mori,
{J. Phys. Soc. Jpn.} {\bf 73}, 116 (2004).

\bibitem{Kobayashi86}
H. Kobayashi, R. Kato, A. Kobayashi, Y. Nishio, K. Kajita, and W. Sasaki,
{Chem. Lett.}  {\bf 1986}, 789 (1986).

\bibitem{Watanabe99}
M. Watanabe, Y. Nogami, K. Oshima, H. Mori, and S. Tanaka, 
{J. Phys. Soc. Jpn.} {\bf 68}, 2654 (1999).

\bibitem{Nogami99}
Y. Nogami, J.\,P. Pouget, M. Watanabe, K. Oshima, H. Mori, S. Tanaka, and T. Mori,
{Synth. Met.} {\bf 103}, 1911 (1999).

\bibitem{Sasaki08}
T. Sasaki, N. Yoneyama, Y. Nakamura, N. Kobayashi, Y. Ikemoto, T. Moriwaki, and H. Kimura,
{Phys. Rev. Lett.} {\bf 101}, 206403 (2008).

\bibitem{Tajima00}
H. Tajima, S. Kyoden, H. Mori, and S. Tanaka,
{Phys. Rev. B} {\bf 62}, 9378 (2000).

\bibitem{Wang03}
N.\,L. Wang, T. Feng, Z.\,J. Chen, and H. Mori,
{Synth. Met.} {\bf 135}, 701 (2003).

\bibitem{Suzuki05}
K. Suzuki, K. Yamamoto, K. Yakushi, and A. Kawamoto,
{J. Phys. Soc. Jpn.} {\bf 74}, 2631 (2005).

\bibitem{Wang01}
N.\,L. Wang, H. Mori, S. Tanaka, J. Dong, and B.\,P. Clayman,
{J. Phys. Condens. Matter} {\bf 13}, 5463 (2001).

\bibitem{Shimano}
R. Shimano (private Communication).

\bibitem{Dressel03}
M. Dressel, N. Drichko, J. Schlueter, and J. Merino,
{Phys. Rev. Lett.} {\bf 90}, 167002 (2003).

\bibitem{Drichko06}
N. Drichko, M. Dressel, C.\,A. Kuntscher, A. Pashkin, A. Greco, J. Merino, and J. Schlueter,
{Phys. Rev. B} {\bf 74}, 235121 (2006).

\bibitem{Kaiser10}
S. Kaiser, M. Dressel, Y. Sun, A. Greco, J.\,A. Schlueter, G.\,L. Gard, and N. Drichko,
{Phys. Rev. Lett.} {\bf 105}, 206402 (2010).

\bibitem{Merino03}
J. Merino, A. Greco, R.\,H. McKenzie, and M. Calandra,
{Phys. Rev. B} {\bf 68}, 245121 (2003).

\bibitem{explanation1}
For instance, in a checkerboard CO, this broad band appears at $3V$ when considering only the nearest-neighbor Coulomb interaction $V$ since the energy cost to move one hole from the perfect checkerboard pattern to its neighboring unoccupied sites is $3V$, which can be reduced in the charge-ordered metallic state owing to the reduction in $V/t$.

\bibitem{Naka13}
M. Naka and H. Seo (in preparation).

\bibitem{Fratini09}
S. Fratini and J. Merino,
{Phys. Rev. B} {\bf 80}, 165110 (2009).

\bibitem{Kagawa}
F. Kagawa (private Communication).

\bibitem{Yoneyama10}
N. Yoneyama, T. Sasaki, N. Kobayashi, K. Furukawa, and T. Nakamura,
{Physica B} {\bf 405}, S244 (2010).

\bibitem{Ono97}
S. Ono, T. Mori, S. Endo, N. Toyota, T. Sasaki, Y. Watanabe, and T. Fukase,
{Physica C} {\bf 290}, 49 (1997).



\end{thebibliography}
\end{document}